\documentclass{article}
\usepackage{spconf,amsmath,amsfonts,graphicx}

\title{Deep attractor network for single-microphone speaker separation }
\name{Zhuo Chen \qquad Yi Luo \qquad Nima Mesgarani}

\address{ Department of Electrical Engineering, Columbia University, New York, NY}
\begin{document}
\ninept
\maketitle
\begin{abstract}
Despite the overwhelming success of deep learning in various speech processing tasks, the problem of separating simultaneous speakers in a mixture remains challenging. Two major difficulties in such systems are the arbitrary source permutation and unknown number of sources in the mixture. We propose a novel deep learning framework for single channel speech separation by creating attractor points in high dimensional embedding space of the acoustic signals which pull together the time-frequency bins corresponding to each source. Attractor points in this study are created by finding the centroids of the sources in the embedding space, which are subsequently used to determine the similarity of each bin in the mixture to each source. The network is then trained to minimize the reconstruction error of each source by optimizing the embeddings. The proposed model is different from prior works in that it implements an end-to-end training, and it does not depend on the number of sources in the mixture. Two strategies are explored in the test time, K-means and fixed attractor points, where the latter requires no post-processing and can be implemented in real-time. We evaluated our system on Wall Street Journal dataset and show 5.49\% improvement over the previous state-of-the-art methods.
\end{abstract}
\begin{keywords}
Source separation, multi-talker, deep clustering, attractor network
\end{keywords}
\section{Introduction}
\label{sec:intro}


Despite the recent advances in deep learning methods for various speech processing tasks such as automatic recognition\cite{asr1,asr2,asr3} and enhancement\cite{se1,se2,se3}, speech separation remains unresolved. Two main difficulties hinder the efficacy of deep learning algorithms to tackle the famous ``cocktail party problem''\cite{Cherry1953}. The first difficulty is referred as the ``permutation problem" \cite{dpcl1}. Most neural network methods are trained to map the input signal to a unique target output which can be a label, a sequence, or regression reference. Permutation problem in speech separation arises due to the fact that the order of targets in the mixture is irrelevant. For example, when separating speakers in mixture of A+B, both (A,B) and (B,A) solutions are acceptable. However, training the neural network with more than one target label per sample will produce conflicting gradients and thus lead to convergence problem, because the order of targets cannot be determined beforehand. For instance, assigning speaker A to the first target position in mixture A+B and mixture A+C will cause confusions when the mixture is B+C, since they both need to be in the second position for  consistency. 

The second problem in using neural network framework for speech separation is the output dimension mismatch problem. Since the number of sources in the mixture can vary, a neural network with fixed number of output nodes does not have the flexibility to separate arbitrary number of sources. 

Two deep learning based methods have been proposed to resolve these problems, which are known as "deep clustering (DC)\cite{dpcl1}" and "permutation invariant training (PIT)\cite{pit}". In deep clustering, a network is trained to generate discriminative embedding for each time-frequency (T-F) bin with points belonging to the same source forced to be closer to each other. DC is able to solve both permutation and output dimension problem to produce the state of the art separation performance. The main drawback of DC is its inefficiency to perform end-to-end mapping, because the objective function is the affinity between the sources in the embedded space and not the separated signals themselves. Minimizing the separation error is done with an unfolding clustering system and a second network, which is trained iteratively and stage by stage to ensure convergence \cite{dpcl2}.  The PIT algorithm solves the permutation problem by pooling over all possible permutations for N mixing sources ($N!$ permutations), and use the permutation with lowest error to update the network.
PIT was first proposed in \cite{dpcl1}, and was later shown to have comparable performance as DC  \cite{pit}. However, PIT approach suffers the output dimension mismatch problem because it assumes a fixed number of sources. PIT also suffers from its computation efficiency, where the prediction window has to be much shorter than context window due to the inconsistency of the permutation both across and within sample segments. 

In this work, we propose a novel deep learning framework which we refer to as the “attractor network” to solve the source separation problem. The term “attractor” is motivated by the well-studied perceptual effects in human speech perception which suggest that the brain circuits create perceptual attractors (magnets) that warp the stimulus space such that to draws the sound that is closest to it, a phenomenon that is called Perceptual Magnet Effect \cite {attractor}. 
Our proposed model works on the same principle by forming a reference point (attractor) for each source in the embedding space which draws all the T-F bins toward itself. Using the similarity between the embedded points and each attractor, a mask is estimated for each sources in the mixture. Since the mask is directly related to the attractor point, the proposed framework can potentially be extended to arbitrary number of sources without the permutation problem. Moreover, the mask learning enables a very efficient end-to-end training scheme and highly reduces the computation complexity compared with DC and PIT.

In Section~\ref{sec:model}, the proposed model is explained and discussed in more detail. In Section~\ref{sec:exp}, we evaluate the performance of proposed system, and the conclusion is drawn in Section~\ref{sec:conclusion}.
\section{Attractor Neural Network}
\label{sec:model}
\subsection{Model}
The neural network is trained to map the mixture sound $X$ to a $k$ dimensional embedding space, such that it minimizes the following objective function: 
\begin{align}
\mathcal{L}&=\sum_{f,t,c}\left \| S_{f,t,c}-X_{f,t} \times M_{f,t,c}\right  \|^2_2
\label{eqn:err}
\end{align}

where $S$ is the clean spectrogram (frequency $F$ $\times$ time $T$) of $C$ sources, $X$ is the mixture spectrogram (frequency $F$ $\times$ time $T$), and $M$ is the mask formed to extract each source. The mask is estimated in the $K$ dimensional embedding space of each T-F bin, represented by $V \in \mathbb R ^{FT \times K}$:
\begin{align}
M_{f,t,c}&=Sigmoid(\sum_k A_{c,k} \times V_{ft,k})
\label{eqn:mask}
\end{align}
where $A \in \mathbb R ^ {C \times K}$ are the attractors for the $C$ sources in the embedding space, learnt during training, which are defined as 
\begin{align}
 A_{c,k}&=\frac{\sum_{f,t} V_{k,ft}\times Y_{c,ft} }{\sum_{f,t} Y_{c,ft}}
 \label{eqn:att}
\end{align}
where $Y \in \mathbb R ^{FT \times C}$ is the source membership function for each T-F bin, i.e., $Y_{tf,c}=1$ if source $c$ has the highest energy at time $t$ and frequency $f$ compare to the other sources.

The objective function in Equation~\ref{eqn:err} consists of three parts. During training, we first compute an embedding $V$ through a forward pass of the neural network for each given mixture. Then an attractor vector is estimated for each source using Equation~\ref{eqn:att}. This can be done in several ways which we will elaborate in Section~\ref{sec:form_att}. The most straightforward method for attractor generation is to find the source centroid, as defined in Equation~\ref{eqn:att}.

Next, we estimate a reconstruction mask for each source by finding the similarity of each T-F bin in the embedding space to each of the attractor vectors $A$, where the similarity metric is defined in Equation~\ref{eqn:mask}. This particular metric uses the inner product followed by a sigmoid function which monotonically scales the masks between $[0,1]$. Intuitively, if an embedding of a T-F bin is closer to one attractor, then it means that it belongs to that source, and the resulting mask for that source will produce larger values for that T-F bin. 
Since the source separation masks for each TF bin should add up to one, particularly in difficult conditions, the sigmoid function in step 2 can replace with softmax function"

\begin{align}
M_{f,t,c}&=Softmax(\sum_k A_{c,k} \times V_{f,t,k})
\label{eqn:mask_softmax}
\end{align}

\begin{figure}[h]
    \centering
    \includegraphics[width=\columnwidth]{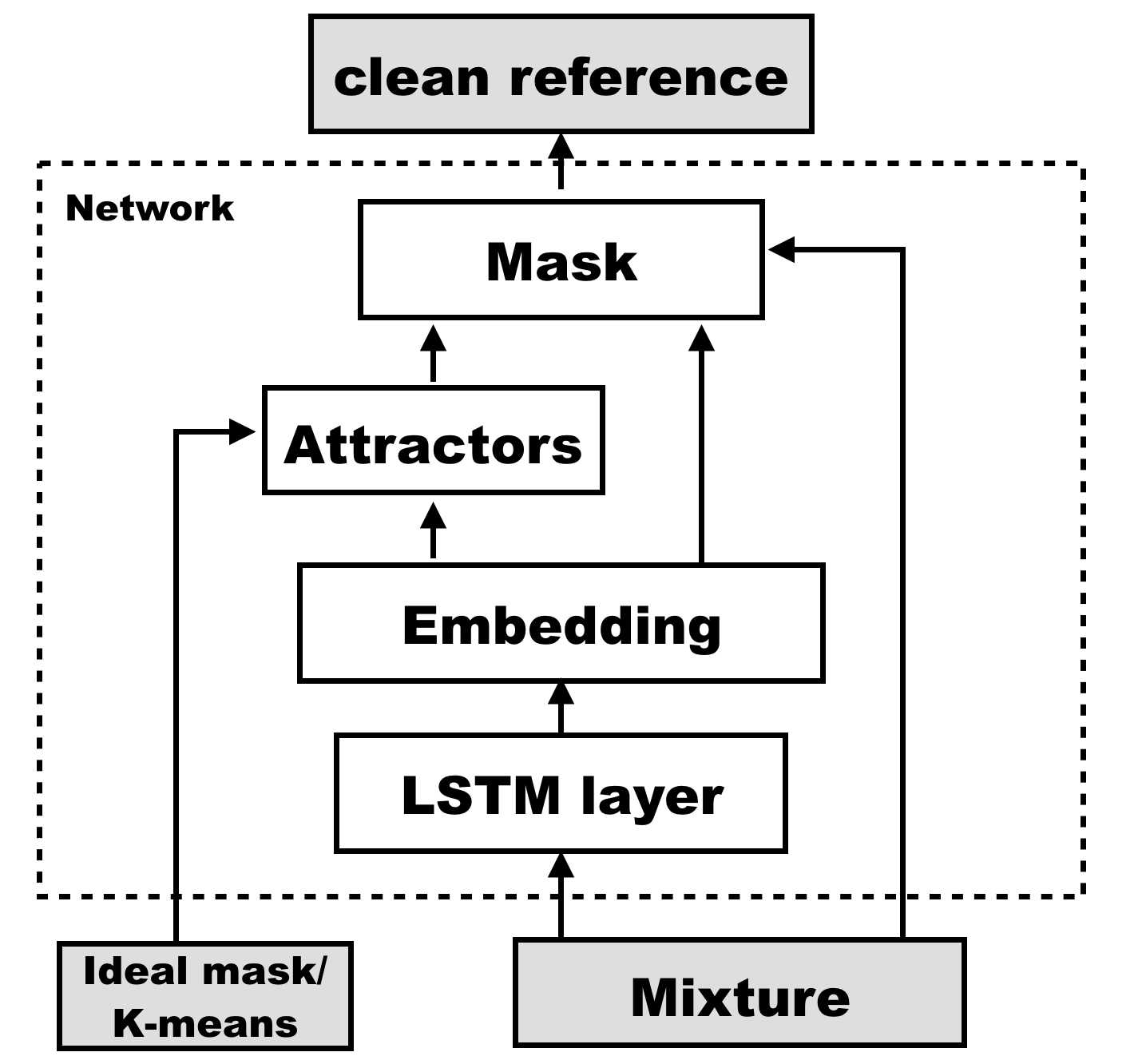}
    \caption{The system architecture. In the training time, a ideal mask is applied to form the attractor, while during the testing time, Kmeans is used to form the attractor. Alternatives for Kmeans is further discussed in Section 2.3}
    \label{fig:sys}
\end{figure}

Finally, a standard $L2$ reconstruction error is used to generate the gradient, as shown in Equation~\ref{eqn:err}. Therefore, the error for each source reflects the difference between the masked signal and the clean reference, forcing the network to optimize the global reconstruction error for better separation. We refer the proposed net as deep attractor network (DANet). Figure \ref{fig:sys} shows the structure of the proposed system.

\subsection{Results}


In comparison with previous methods, DANet network has two distinct advantages. Firstly, DANet removes the stepwise pre-training required in DC method to enable end-to-end training. Another advantage of DANet arises from flexibility in the number of sources in the mixture.

\subsection{Estimation of attractor points}
\label{sec:form_att}
Attractor points can be estimated using various methods other than the average used in Equation~\ref{eqn:att}. One possibility is to use weighted average. Since the attractors represents the source center of gravity, we can include only the embeddings of the most salient T-F bins, which leads to more robust estimation. We investigate this strategy by using an amplitude threshold in the estimation of the attractor. Alternatively, a neural network model may also be used to pick the representative embedding for each source, an idea which shares similarities with encoder-decoder attention networks \cite{attention1,attention2}. 

\begin{figure*}[t]
    \centering
    \includegraphics[width=17.5cm]{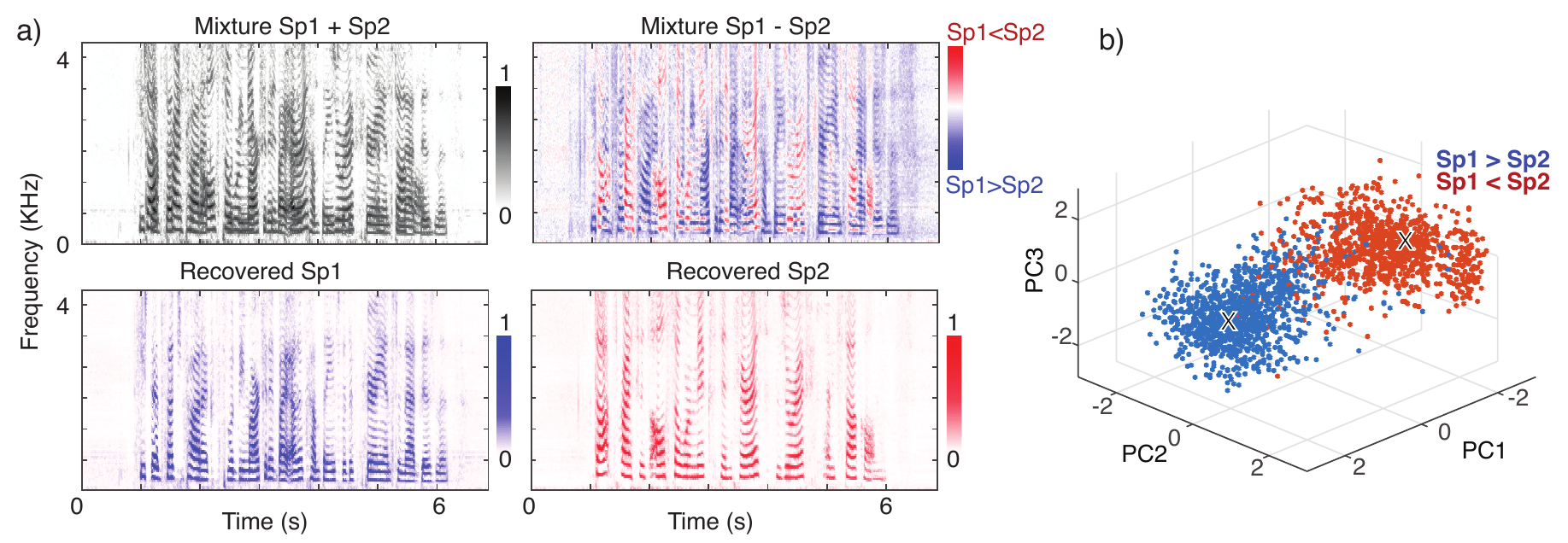}
    \caption{Location of T-F bins in the embedded space. Each dot visualizes the first three principle components of one T-F bin, where colors distinguish the relative power of speakers, and the location of attractors is marked with X.}
    \vspace{0.1cm}
    \label{fig:attractor_1}
\end{figure*}

During test time, because the true assignment $Y$ is unknown, we incorporate two strategies to form the attractor points. The first is similar to the strategy used in DC, where the centers are found using post K-means algorithm. The second method is based on the observation that the location of the attractors in the embedding space is relatively stable. This observation is shown in Figure~\ref{fig:attractor_1}, where each pair of dots corresponds to the attractor found for the two speakers in a given mixture. Figure~\ref{fig:attractor_1} shows two principle stable attractor pairs for all the mixtures used, however, this observation needs to be tested in more depth and different tasks and datasets.  

\subsection{Relation with DC and PIT}
\label{sec:relation}
The objective function of DC is shown in Equation~\ref{eqn:DC}, where $Y$ is the indicator function which is equivalent to a binary mask, and $V$ is the learnt embedding:
\begin{align}
\mathcal{L}&=\left \| YY^T- VV^T \right  \|^2_2
\label{eqn:DC}
\end{align}

Since $Y$ is orthogonal and constant for each mixture, by multiplying $Y^T$ and a normalizer $U=(Y^T Y)^{-1}$ to both term, we can get an objective function that is a special case of the attractor network, as in Equation~\ref{eqn:DC2}:
\begin{align}
\mathcal{L}&=\left \|  Y^T- UY^TVV^T \right  \|^2_2
\label{eqn:DC2}
\end{align}

In Equation~\ref{eqn:DC2}, $UY^TV$ can be viewed as an averaging step, where the embeddings are summed according to the label, and the resulted center is multiplied with the embedding matrix $V$ to measure the similarity between each embedding and the center, and compared with the ground truth binary mask. When the learnt $V$ is optimum, i.e, $VV^T=YY^T$, equation ~\ref{eqn:DC} and ~\ref{eqn:DC2} are equivalent.



On the other hand, when the attractor vectors are considered as free parameters in the network, DANet reduces to a classification network \cite{se1,se2}, and Equation~\ref{eqn:err} becomes a fully-connected layer. In this case, PIT becomes necessary since the mask has no information about the source and the problem of fixed output dimension arises. In contrast, the freedom of the network to form attractor points during the training allows the system to use the affinity between samples where no constraint is on the number of patterns found, therefore allowing the network to be independent of the number of sources. 
The flexibility of the network in choosing the attractor points is helpful even in two-source separation problem, because the two sources may have very different structures. As can be seen in Figure~\ref{fig:attractor_all}, our proposed method trained in speaker separation tasks has ended up finding 2 attractor pairs (4 points in the embedding space), which can be expected to increase in harder problems. 

\section{Evaluation}
\label{sec:exp}
\subsection{Experimental setup}

We evaluated our proposed model on the task of simultaneous speech separation.
We used the corpus introduced in \cite{dpcl1}, which contained a 30 h training set and a 10 h validation set generated by randomly selecting utterances from different speakers in the Wall Street Journal (WSJ0) training set si\_tr\_s, and mixing them at various signal-to-noise ratios (SNR) randomly chosen between 0 dB and 10 dB. 
5 h evaluation set was generated similarly as above, using utterances from 16 unseen speakers from si\_dt\_05 and si\_et\_05 in WSJ0 dataset. 
Additionally, we constructed a three speaker mixture dataset for three speaker separation evaluation from same WSJ set, which had 30h training, 10 hours validation and 5 hours testing data, with mixing SNR at -5$\sim$ 5 dB. We ensured that in each three speaker mixture, there existed both female and male speakers.
All data were resampled to 8 kHz to reduce computational and memory costs. The log spectral magnitude was served as input feature, computed using short-time Fourier transform (STFT) with 32 ms window length, 8 ms hop size, and the square root of hanning window.



The network contained 4 Bi-directional LSTM \cite{lstm} layers with 600 hidden units in each layer. The embedding dimension was set to 20, resulting in a fully-connected feed-forward layer of 2580 hidden units (20 $\times$ 129) after the BLSTM layers. We split the input features into non-overlapping chunks of 100-frame length as the input to the network. RMSprop algorithm \cite{rmsprop} was used for training with an exponential learning rate decaying strategy, where the learning rate started at $10^{-4}$ and ends at $3\times10^{-6}$. The total number of epochs was set to be 150, and we used the cost function in Equation~\ref{eqn:err} on the validation set for early stopping. The criteria for early stopping is no decrease in the loss function on validation set for 10 epochs. We constructed a Deep Clustering (DC) network with the same configuration which is used as the baseline. 

We report the results in terms of signal-to-distortion ratio (SDR, which we define as scale-invariant SNR here)\cite{dpcl2}, signal-to-artifacts ratio (SAR), and signal-to-interference ratio (SIR). The results are shown in Table 1. 

\subsection{Separation examples}
Figure~\ref{fig:attractor_1} shows an example of mixture, the difference between the two speakers in the mixture, and the separated spectrograms of the two speakers using DANet. Also visualized in Figure~\ref{fig:attractor_1} is the embeddings of the mixture projected onto its first Principal Components. In Figure~\ref{fig:attractor_1}, each point represents one T-F bin in the embedding space. Red and blue dots correspond to the T-F bins where speaker one or two have greater energy accordingly. The location of attractors are marked by x. It shows that two symmetric attractor centers are formed, each corresponding to one of the speakers in the mixture. A clear boundary can be observed in the figure, showing that the network successfully pulled the two speakers apart toward their corresponding attractor points. 

\begin{figure}[h]
    \centering
    \includegraphics[width=\columnwidth]{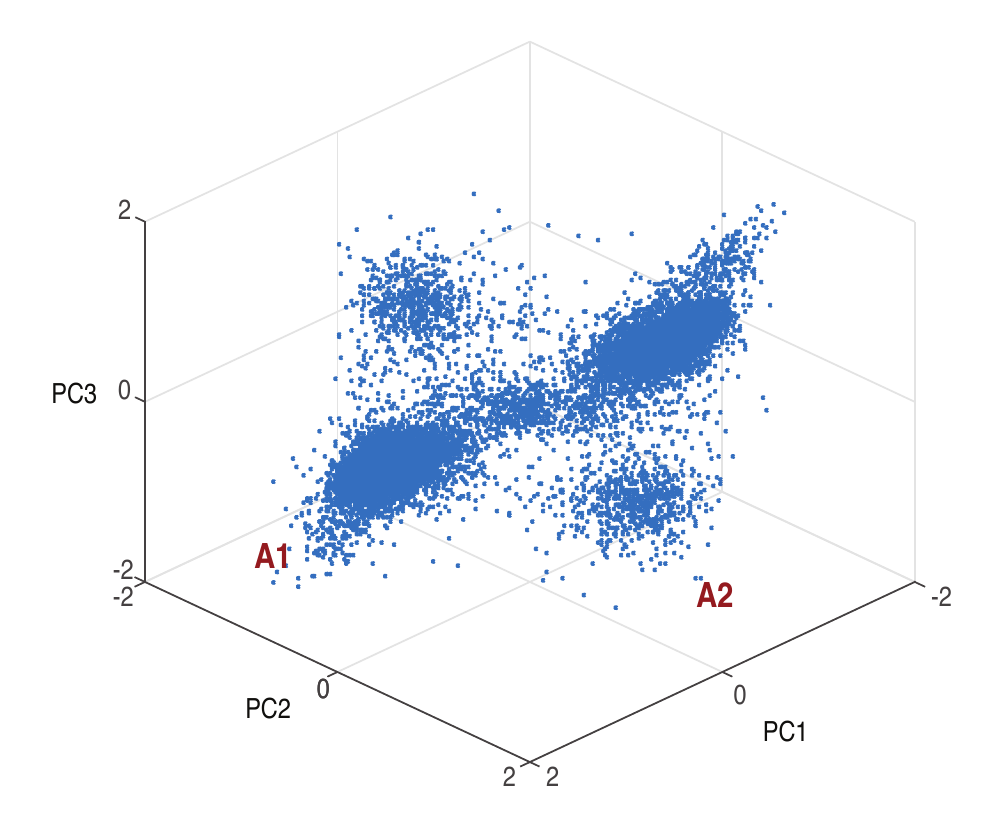}
    \caption{Location of attractor points in the embedding space. Each dot corresponds to one of the 10000 mixtures sounds, visualized using the first three principal components. Two distinct attractor pairs are visible (denoted by A1 and A2).}
    \label{fig:attractor_all}
\end{figure}

Figure~\ref{fig:attractor_all} shows the location of attractors for 10,000 mixture examples, mapped onto the 3-dimensional space for visualization purpose using Principal Component Analysis. It suggests that the network may have learned two attractor pairs (4 symmetric centers), marked by A1 and A2. This observation confirms our intuition of the DANet mentioned in Section ~\ref{sec:relation}, that DANet has the ability to discover different number of attractors in an unsupervised way, and therefore, form complex separation strategies. Although the task considered in this study is already challenging, one can imagine much more difficult separation scenarios, where the number of speakers in the mixture is large and can change over time. The ability of DANet to form new attractors may prove to be crucial in such cases, because any effort in pre-setting the number of mixing patterns, as done in methods such as PIT, will hinder the generalization ability of the network. Figure~\ref{fig:attractor_all} also suggests that hierarchical clustering methods can be more suitable, where attractors can drive a hierarchical grouping of sources, allowing a better representation of audio signals. We will explore these issues in future work.

\subsection{Results}

Table 1 shows the evaluation results for different networks (example sounds can be found here \cite{website}). 
Although the plain DANet already outperforms the DC baseline, adding a simple threshold on T-F samples included in the formation of the attractor yields further improved performance, presumably due to the increased influence of salient segments. On the other hand, the performance boost suggests that better attractor formation procedures can be utilized to estimate the attractors, such as joint optimizing of the network parameters. Moreover, by applying curriculum training strategy \cite{dpcl2}, which we continue training the network with 400-frame length input, DANet achieves the best overall performance. 

In the last experiment in Table 1, a fixed pair of attention vector collected from the training data is used, corresponding to the A1 pair in Figure~\ref{fig:attractor_all}. This pre-set attractor is able to generalize well to the unseen mixtures and produced high quality separation, however it was slightly worse than the best model. Compared with K-means, this has the advantage that it can be easily implemented in real-time using a frame-by-frame pipeline. Based on this observation, when more attractors are required (e.g. in more complex tasks), a collection of attractor codebook and a simple classifier could be implemented for real-time processing. 

In three speaker separation experiment, shown in Table 2, the proposed system significantly outperforms the deep clustering baseline. This result maybe expected since deep clustering was trained to estimate binary mask, while the deep attractor network focuses on the signal reconstruction. When the mixture is relatively simple, the binary mask could generate high quality separation. However, for more complex mixtures, or when one source is significantly louder than the other, the binary mask usually lead to large bias to the loudest source, and thus result in unsatisfiable separation for weaker source. Note that in the three speaker experiment, the network was trained using softmax objective as shown in (\ref{eqn:mask_softmax}).


\section{Conclusion}
\label{sec:conclusion}
In this work, we proposed a novel neural network framework called deep attractor network for general source separation problem. The network forms attractor points in a high-dimensional embedding space of the signal, and the similarity between attractors and time-frequency embeddings are then converted into a soft separation mask. We showed that the proposed framework can perform end-to-end, real-time separation, can work on different number of mixing sources, and is more general comparing to deep clustering and classification based approaches. The experiment on two speaker separation task confirmed its efficacy and potential for extension to general source separation problems. 

\begin{table}[!htbp]
\label{tab:res}
\centering
\begin{tabular}{c|ccc}
\hline
 & GNSDR & GSAR & GSIR \\
 \hline
DC & 9.1 & 9.5 & \textbf{22.2} \\
\hline
DANet & 9.4 & 10.1 & 18.8 \\
DANet-50\% & 9.4 & 10.4 & 17.3 \\
DANet-70\% & 9.6 & 10.3 & 18.7 \\
DANet-90\% & 9.6 & 10.4 & 18.1 \\
DANet-90\%$^\ddagger$ & \textbf{10.5} & \textbf{11.1} & 20.3 \\
fix-DANet-90\% & 9.5 & 10.4 & 17.8 \\
\hline
\end{tabular}
\caption{Evaluation metrics for networks with different configurations. The percentage suffixes stand for the salient weight threshold used during training. $^\ddagger$: curriculum training with 400-frame length input\cite{dpcl2}. fix-DANet: fixed attractor points calculated by training set are used.}
\end{table}

\begin{table}[!htbp]
\label{tab:res_3spk}
\centering
\begin{tabular}{c|ccc}
\hline
 & GNSDR & GSAR & GSIR \\
 \hline
DC & 6.3 & 2.3 & 12.6 \\
DANet & 7.7 & 3.9 & 13.2 \\
DANet$^\ddagger$ & \textbf{8.8} & \textbf{5.0} & \textbf{15.0} \\
\hline
\end{tabular}
\caption{Evaluation results for three speaker separation. $^\ddagger$: curriculum training with 400-frame length input.}
\end{table}

\section{Acknowledgement}
The authors would like to thank Drs. John Hershey and Jonathan Le Roux of Mitsubishi Electric Research Lab for constructive discussions. This work was funded by a grant from National Institute of Health, NIDCD, DC014279, National Science Foundation CAREER Award, and the Pew Charitable Trusts.

\vfill\pagebreak

\bibliographystyle{IEEEbib}
\bibliography{Template}

\end{document}